\documentclass[12pt]{article}
\usepackage{epsfig}
\usepackage{amssymb}
\usepackage{amsmath}
\usepackage{amsfonts}
\usepackage{graphicx}
\usepackage{mathrsfs}
\usepackage[dvips]{color}
\usepackage{multirow}

\newcommand{\R}{\mathbb{R}}
\newcommand{\C}{\mathbb{C}}

\newcommand{\fn}{{\mathfrak{n}}}

\newcommand{\fs}{\mathfrak{s}}

\newcommand{\fz}{\mathfrak{z}}

\newcommand{\fK}{\mathfrak{K}}

\newcommand{\bH}{\mathbf{H}}
\newcommand{\bI}{\mathbf{I}}

\newcommand{\bM}{\mathbf{M}}
\newcommand{\bcM}{\boldsymbol{\cM}}

\newcommand{\cF}{\mathcal{F}}

\newcommand{\cK}{\mathcal{K}}

\newcommand{\cM}{\mathcal{M}}

\newcommand{\cO}{\mathcal{O}}

\newcommand{\cR}{\mathcal{R}}

\newcommand{\be}{\begin{equation}}
\newcommand{\ee}{\end{equation}}
\newcommand{\bea}{\begin{eqnarray}}
\newcommand{\eea}{\end{eqnarray}}
\newcommand{\nn}{\nonumber}

\newcommand{\ed}{\end{document}}

\newcommand{\bi}{\begin{itemize}}
\newcommand{\ei}{\end{itemize}}

\newcommand{\bce}{\begin{center}}
\newcommand{\ece}{\end{center}}

\newcommand{\sT}{\mathscr{T}}

\newcommand{\RE}{{\rm Re}}
\newcommand{\IM}{{\rm Im}}

\begin{document}

\title{Solving scattering problems in the half-line using methods developed for scattering in the full line}

\author{Ali~Mostafazadeh\thanks{E-mail address: amostafazadeh@ku.edu.tr}~
\\ Departments of Mathematics and Physics,
Ko\c{c} University,\\ 34450 Sar{\i}yer,
Istanbul, Turkey}

\date{ }
\maketitle

\begin{abstract}
We reduce the solution of the scattering problem defined on the half-line $[0,\infty)$ by a real or complex potential $v(x)$ and a general homogenous boundary condition at $x=0$ to that of the extension of $v(x)$ to the full line that vanishes for $x<0$. We find an explicit expression for the reflection amplitude of the former problem in terms of the reflection and transmission amplitudes of the latter. We obtain a set of conditions on these amplitudes under which the potential in the half-line develops bound states, spectral singularities, and time-reversed spectral singularities where the potential acts as a perfect absorber. We examine the application of these results in the study of the scattering properties of a $\delta$-function potential and a finite barrier potential defined in $[0,\infty)$, discuss optical systems modeled by these potentials, and explore the configurations in which these systems act as a laser or perfect absorber. In particular, we derive an explicit formula for the laser threshold condition for a slab laser with a single mirror and establish the surprising fact that a nearly perfect mirror gives rise to a lower threshold gain than a perfect mirror. We also offer a nonlinear extension of our approach which allows for utilizing a recently developed nonlinear transfer matrix method in the full line to deal with finite-range nonlinear scattering interactions defined in the half-line. 




\end{abstract}

\section{Introduction}

The study of scalar waves propagating in the half-line, $[0,\infty)$, has a long history \cite{dealfaro-regge}. This is mostly because of the essential role it plays in performing scattering calculations for spherically symmetric systems  \cite{chew}. More importantly, many of the developments in scattering and inverse scattering theories have their root in this subject \cite{IS}. Interactions that cause the scattering of a wave in the half-line are of two types, those  that affect the wave as it propagates throughout the interior of the half-line, i.e., $(0,\infty)$, and those corresponding to the response of the boundary point $x=0$. This is in contrast with the scattering interactions in the full line, $\R$, which has no boundary. The purpose of the present article is to make a link between the scattering problems in the half-line and the full line with the intention of using the known results for the solution of the latter to address the former.

Consider a time-harmonic scalar wave $e^{-i\omega t}\psi(x)$ with $\psi(x)$ satisfying the Schr\"odinger equation,
    \be
    -\psi''(x)+v(x)\psi(x)=k^2\psi(x),
    \label{sch-eq}
    \ee
in the half-line, where $v:[0,\infty)\to\C$ is a piecewise continuous scattering potential\footnote{By being a scattering potential we mean that every solution of the Schr\"odinger equation (\ref{sch-eq}) tends to a linear combination of plane waves as $x\to\infty$. This is the case for potentials $v(x)$ satisfying the Faddeev condition $\int_0^\infty (1+x)|v(x)|dx<\infty$, \cite{yafaev}.}, and $k$ is the wavenumber of the incident wave. Our main purpose is to study the scattering problems defined by (\ref{sch-eq}) and the boundary condition
    \be
    \alpha\,\psi(0)+k^{-1}\beta\,\psi'(0)=0,
    \label{bc}
    \ee
where $\alpha$ and $\beta$ are possibly $k$-dependent real or complex parameters fulfilling $|\alpha|^2+|\beta|^2\neq 0$. Note that the Dirichlet, Neumann, and Robin boundary conditions are special cases of (\ref{bc}) that correspond to $\alpha\neq 0 =\beta$, $\alpha=0\neq\beta$, and $\alpha\neq 0=\beta-k$, respectively.

To motivate the choice of the boundary condition (\ref{bc}), we consider a time-harmonic wave $e^{-i\omega t}\Psi(x)$ propagating in the full line ($\R$) with a source located at $x=+\infty$. Suppose that $\Psi(x)$ solves the time-independent Schr\"odinger equation for a general (piecewise continuous) extension of the potential $v(x)$ to $\R$, namely
    \[V(x):=\left\{\begin{array}{ccc}
    v_-(x) & {\rm for} & x<0,\\
    v(x) & {\rm for} & x\geq 0,
    \end{array}\right.\]
where $v_-:(-\infty,0)\to\C$ is an unspecified potential defined in $(-\infty,0)$. Let us explore conditions under which
    \be
    \Psi(x)=\psi(x)~~{\rm for}~~x\in[0,\infty).
    \label{condi-psi}
    \ee
    
First, imagine that $v(x)$ takes a constant value $\nu$ in an open interval of the form $(0,\epsilon)$. Then in view of (\ref{condi-psi}), it is easy to see that for $x\in (0,\epsilon)$,
    \be
    \psi(x)=\Psi(x)=\left\{
    \begin{array}{ccc}
    A_0 e^{i\tilde kx}+B_0 e^{-i\tilde kx} &{\rm for}& k^2\neq\nu,\\
    A_0+B_0 kx &{\rm for}& k^2=\nu,
    \end{array}\right.
    \label{psi-zero}
    \ee
where $A_0$ and $B_0$ are complex coefficients, and $\tilde k:=\sqrt{k^2-\nu}$. Now, introduce
    \be
    \cR_-:=\frac{A_0}{B_0},
    \label{ref-zero}
    \ee
which for real and positive values of $\tilde k$ represents the reflection amplitude of the potential
    \[V_-(x):=\left\{\begin{array}{ccc}
    v_-(x) & {\rm for} & x< 0,\\
    0 & {\rm for} & x\geq 0,\end{array}\right.\]
for a right-incident plane wave with wavenumber $\tilde k$. Differentiating (\ref{psi-zero}) and expressing $\psi(0):=\lim_{x\to 0^+}\psi(x)$ and $\psi'(0):=\lim_{x\to 0^+}\psi'(x)$ in terms of $B_0$ and $\cR_-$, we find
    \be
    \begin{array}{ccc}
    i(\cR_--1)\psi(0)+\tilde k^{-1}(\cR_-+1)\psi'(0)=0 & {\rm for} & k^2\neq\nu,\\[6pt]
    \psi(0)-k^{-1}\cR_-\psi'(0)=0 &{\rm for} &k^2=\nu.
    \end{array}
    \label{bc-zero}
    \ee
This shows that in order to determine the behavior of the wave in $[0,\infty)$ we actually do not need a detailed knowledge of $v_-(x)$; the information about $\cR_-$ suffices for this purpose. The argument leading to this conclusion holds generally, for we can make $\epsilon$ arbitrarily small. The fact that the boundary conditions given by (\ref{bc-zero}) are special cases of (\ref{bc}) provides the main motivation for the study of the scattering problems defined by the Schr\"odinger equation (\ref{sch-eq}) and the boundary conditions (\ref{bc}) in the half-line.

Because $v(x)$ is a scattering potential, for $x\to\infty$ it decays to zero at such a rate that every solution $\psi(x)$ of (\ref{sch-eq}) satisfies
    \be
    \psi(x)\to A_+e^{ikx}+B_+e^{-ikx}~~~{\rm for}~~~x\to\infty,
    \label{bc-p}
    \ee
where $A_+$ and $B_+$ are possibly $k$-dependent complex coefficients. The solution of the scattering problem for $v(x)$ means finding its reflection amplitude,
    \be
    \cR:=\frac{A_+}{B_+}.
    \label{ref-coeff}
    \ee

If for a real and positive value of $k$, $A_+=0\neq B_+$, the reflection amplitude vanishes, and the incident wave is completely absorbed by the system, i.e., it acts as a perfect absorber. Another interesting situation is when the converse occurs. In this case the system emits a purely outgoing plane wave, i.e.,  $A_+\neq 0=B_+$, $\cR$ blows up for a real and positive value of $k$, and $k^2$ is called a spectral singularity. This is a mathematical concept introduced by Naimark in 1954 \cite{naimark} and further developed and studied by other mathematicians \cite{kemp,schwartz,lyantse,gasimov-1968,gasimov-maksudov,langer,nagy,bairamov,guseinov}. Surprisingly, the physical significance of this concept was revealed more than half a century after its discovery \cite{prl-2009}. This led to a surge of research activity in the study of physical aspects of spectral singularities \cite{Longhi,CP-2012,GC-2013-14,prl-2013,HR,Li-2014,Zhang-2015,wang-2016,Hang-2016,Kalozoumis-2016,Pendharker-2016,Konotop-ol-2017,Zhang-2017,Ahmed-2018,Jin-2018,Konotop-2018,Midya-2018,Zezyulin-2018,Muller-2018,Li-2019,Konotop-2019} and unraveled their intimate relation to the physics of lasers; the mathematical condition for the emergence of a spectral singularity for optical potentials modeling the scattering of electromagnetic waves coincides with the laser threshold condition which marks the initiation of laser oscillations \cite{pra-2011a,spherical,WGM}. See also \cite{liu,reddy,ap-2016,jo-2017,ap-2018}.

It is not difficult to see that the perfect absorption and emission phenomena corresponding to the conditions $\cR=0$ and $\cR=\infty$, are related to one another by time-reversal transformation \cite{book-chapter}. This has led to the use of the term ``anti-laser'' for optical perfect absorbers \cite{antilaser1,longhi-2010,antilaser2,gupta-2012,sun-2014,fan-2014,wong-2016,baranov-2017,lanoy-2018,mustafa-2018-9}.

A more common situation is when $A_+\neq 0$, $B_+= 0$, and consequently $\cR=\infty$ for a complex value of $k$ with a positive imaginary part. In this case, the corresponding solution of the Schr\"odinger equation decays exponentially as $x\to\infty$. Therefore, it is square-integrable. This shows that $k^2$ is a genuine eigenvalue of the Schr\"odinger operator $H:=-\partial_x^2+v(x)$. If such a value of $k$ is purely imaginary, $k^2$ is a negative real number. This implies that the norm of the corresponding eigenfunction $\psi_k$ does not decay in time;
    \[ \parallel e^{-i t H}\psi_k \parallel=\parallel e^{-i t k^2}\psi_k \parallel
    =\parallel \psi_k\parallel,\]
where $e^{-itH}$ is the time-evolution operator. Therefore, $\psi_k$ identifies a bound state of the potential. If $k$ is not purely imaginary, $k^2$ develops an imaginary part, and the norm of $\psi_k$ decays or grows exponentially in time depending on the sign of $\IM(k^2)$; these correspond to the resonances and anti-resonances of the system, respectively.

\section{Scattering in half-line from the full line}

Consider the trivial extension of the potential $v(x)$ to the whole real line that  vanishes for $x<0$, i.e.,
    \be
    V_+(x):=\left\{\begin{array}{ccc}
    0 &{\rm for} & x<0,\\
    v(x) &{\rm for} & x\geq 0.\end{array}\right.
    \label{V+}
    \ee
Clearly, every solution $\psi(x)$ of the Schr\"odinger equation (\ref{sch-eq}) for the potential $V_+(x)$ satisfies (\ref{bc-p}) and
    \be
    \psi(x)=A_-e^{ikx}+B_-e^{-ikx}~~~{\rm for}~x\leq 0,
    \label{bc-m}
    \ee
where $A_-$ and $B_-$ are possibly $k$-dependent complex coefficients. The restriction of these solutions to the half-line $[0,\infty)$ gives the scattering solutions defined by (\ref{sch-eq}) and (\ref{bc}), if we choose $A_-$ and $B_-$ in such a way that $\psi(0)$ and  $\psi'(0)$ satisfy (\ref{bc}). In view of (\ref{bc-m}), this gives
    \be
    B_-=-\gamma A_-
    \label{bc-2}
    \ee
where
    \be
    \gamma:=\frac{\alpha+i\beta}{\alpha-i\beta}.
    \label{gamma=}
    \ee
Notice that the Dirichlet and Neumann boundary conditions respectively correspond to $\gamma=1$ and $\gamma=-1$.

Now, consider the transfer matrix of the extended potential $V_+(x)$, \cite{book-chapter,sanchez-soto}. This is a $2\times 2$ complex matrix $\bM$ with unit determinant that by definition satisfies
    \be
    \bM\left[\begin{array}{c}
    A_-\\
    B_-\end{array}\right]=\left[\begin{array}{c}
    A_+\\
    B_+\end{array}\right].
    \label{M-def}
    \ee
The scattering properties of $V_+(x)$ are encoded in its left/right reflection ($R^{l/r}$) and transmission ($T^{l/r}$) amplitudes \cite{muga}. These are defined for the left- (respectively right-) incident waves corresponding to $B_+=0$ (respectively $A_-=0$) according to the following relations.
    \be
    \begin{aligned}
    &R^l:=\frac{B_-}{A_-}, &&\hspace{1cm} T^l:=\frac{A_+}{A_-}\\
    &R^r:=\frac{A_-}{B_+}, &&\hspace{1cm} T^r:=\frac{B_-}{B_+}.
    \end{aligned}
    \label{RT-def}
    \ee
A key observation signifying the importance of the transfer matrix is that its entries determine the reflection and transmission amplitudes of the potential \cite{book-chapter,sanchez-soto}. Specifically,
	\be
    \begin{aligned}
    &R^l=-\frac{M_{21}}{M_{22}},\hspace{1cm}
    &&T^l=\frac{\det(\bM)}{M_{22}},\\
    &R^r=\frac{M_{12}}{M_{22}},\hspace{1cm}
    &&T^r=\frac{1}{M_{22}}.
    \end{aligned}
    \label{RRT=}
    	\ee
These equations together with the fact that $\det(\bM)=1$ imply transmission reciprocity, $T^l=T^l$, which allows us to drop the superscript $l/r$ in $T^{l/r}$. They also lead to 
    \begin{align}
    &M_{11}=T-\frac{R^lR^r}{T},
    && M_{12}=\frac{R^r}{T},
    &&M_{21}=-\frac{R^l}{T},
    &&M_{22}=\frac{1}{T}.
    \label{Mij=0}
    \end{align}

Solving (\ref{ref-coeff}) for $A_+$ and substituting the result together with (\ref{bc-2}) in (\ref{M-def}), we obtain
    \be
    A_-\bM\left[\begin{array}{c}
    1\\
    -\gamma\end{array}\right]=B_+\left[\begin{array}{c}
    \cR\\
    1\end{array}\right].
    \ee
This is equivalent to
    \bea
    &&(M_{11}-\gamma M_{12})A_-=\cR B_+,
    \label{q101}\\
    &&(M_{21}-\gamma M_{22})A_-=B_+.
    \label{q102}
    \eea
Substituting the second of these equations in the first and making use of (\ref{Mij=0}), we find
    \bea
    \cR&=&\frac{M_{11}-\gamma M_{12}}{M_{21}-\gamma M_{22}}
    \label{ref-coeff1}\\
    &=&\frac{T^2}{R^r-\gamma}-R^l.
    \label{ref-coeff2}
    \eea
According to (\ref{ref-coeff1}) and (\ref{ref-coeff2}), we can solve the original scattering problem, which is given in the half-line, by solving the scattering problem defined by the extended potential $V_+(x)$ in the full line. 

Next, we recall that the bound states, resonances, and spectral singularities of the original scattering problem are given by the singularities of the reflection amplitude $\cR$. Because $\det\bM=1$, the numerator and denominator of the right-hand side of (\ref{ref-coeff1}) cannot vanish simultaneously. This implies that the bound states, resonances, and spectral singularities correspond to (real or complex) values of $k$ where $M_{21}=\gamma M_{22}$ or equivalently
    \be
    R^r=\gamma.
    \label{ss}
    \ee
Similarly, the original potential that is defined on the half-line acts as a perfect absorber for the real and positive values of $k$ ensuring $M_{11}=\gamma M_{12}$. In view of (\ref{RRT=}), we can alternatively express this condition in the form
    \be
    T^2-R^l(R^r-\gamma)=0.
    \label{ss}
    \ee
Furthermore, because $T$ can never vanish, (\ref{ss}) implies $R^l\neq 0$. As a result, (\ref{ss}) means
    \be
    R^r=\frac{T^2}{R^l}+\gamma.
    \label{ss1}
    \ee

\section{$\delta$-function potential in the half-line}

Consider the scattering problem defined by (\ref{sch-eq}) and (\ref{bc}) for the $\delta$-function potential,
    \be
    v(x)=\fz\,\delta(x-a),~~~~~~x\geq 0,
    \label{delta-half}
    \ee
where $\fz$ is a real or complex coupling constant, and $a$ is a positive real parameter. The choice of $\alpha=1$ and $\beta=0$ (and hence $\gamma=1$) corresponds to the Dirichlet boundary condition which we encounter in dealing with the potential
    \be
    v(x)=\left\{\begin{array}{ccc}
    +\infty & {\rm for} & x<0,\\
    \fz\,\delta(x-a) & {\rm for} & x\in [0,\infty).
    \end{array}\right.
    \nn
    \ee
Our aim is to solve the scattering problem for (\ref{delta-half}) using the the information about the transfer matrix of its trivial extension to $\R$, namely
    \be
    V_+(x)=\fz\,\delta(x-a),~~~~~~x\in\R.
    \label{delta}
    \ee

The quickest method of computing the transfer matrix for the $\delta$-function potential (\ref{delta}) is to make use of the fact that it can be expressed as the time-ordered exponential of a non-Hermitian two-level effective Hamiltonian \cite{ap-2014-pra-2014a};
    \bea
    \bM&=&\sT\exp\left\{-i\int_{-\infty}^\infty dx \bH(x)\right\}\nn\\
    &=&\bI+\sum_{n=1}^\infty(-i)^n\int_{-\infty}^\infty dx_n
    \int_{-\infty}^{x_n} dx_{n-1}\cdots\int_{-\infty}^{x_2}dx_1
    \bH(x_n) \bH(x_{n-1})\cdots\bH(x_1),
    \label{M=}
    \eea
where $\sT$ is the time-ordering operator with $x$ playing the role of time, $\bI$ is the $2\times 2$ identity matrix, and $\bH(x)$ is the matrix Hamiltonian defined by
    \be
    \bH(x):=\frac{V_+(x)}{2k}\left[\begin{array}{cc}
    1 & e^{-2ikx}\\
    - e^{2ikx} & -1\end{array}\right].
    \label{H=}
    \ee
For the $\delta$-function potential (\ref{delta}), (\ref{H=}) gives
    \be
    \bH(x)=\tilde\fz\,\delta(x-a)\boldsymbol{\cK}_a,
    \label{H=2}
    \ee
where
    \begin{align}
    &\tilde\fz:=\frac{\fz}{2k},
    &&\boldsymbol{\cK}_a:=\left[\begin{array}{cc}
    1 & e^{-2ika}\\
    - e^{2ika} & -1\end{array}\right].
    \label{K-def}
    \end{align}
Substituting (\ref{H=2}) in (\ref{M=}) and noting that $\boldsymbol{\cK}_a^2$ is the null matrix, we find
    \be
    \bM=e^{-i\tilde\fz\boldsymbol{\cK}_a}=\bI-i\tilde\fz\,\boldsymbol{\cK}_a.
    \label{M-delta=}
    \ee
Therefore,    
	\begin{align}
    &M_{11}=1-i\tilde\fz,
    &&M_{12}=-i\tilde\fz e^{-2ika},
    &&M_{21}=i\tilde\fz e^{2ika},
    &&M_{22}=1+i\tilde\fz.
    \label{Mij=}
    \end{align}
In view of (\ref{RRT=}), these relations imply
    \begin{align}
    &R^l=-\frac{i\tilde\fz e^{2ika}}{1+i\tilde\fz},
    && R^r=-\frac{i\tilde\fz e^{-2ika}}{1+i\tilde\fz},
    &&T=\frac{1}{1+i\tilde\fz}.
    \end{align}

If we substitute (\ref{Mij=}) in (\ref{ref-coeff1}), we obtain the following expression for the reflection amplitude of the $\delta$-function potential on the half-line (\ref{delta-half}) with the boundary condition at $x=0$ given by (\ref{bc}).
    \be
    \cR=
    \frac{-2k+i\fz(1-\gamma e^{-2ika})}{2\gamma k+i\fz(\gamma-e^{2ika})}.
    \label{R=delta}
    \ee
Because the bound states, resonances, and spectral singularities correspond to the singularities of $\cR$ in the complex $k$-plane, these are given by values of $k$ for which
    \be
    2\gamma k+i\fz(\gamma-e^{2ika})=0.
    \label{zeros}
    \ee
The system acts as a perfect absorber for the incident plane waves whose wavenumber $k$ fulfills
    \be
    2k+i\fz(\gamma e^{-2ika}-1)=0.
    \label{CPA}
    \ee

\subsection{Bound states}

Bound states correspond to solutions of (\ref{sch-eq}) for the positive imaginary values of $k$ that satisfy (\ref{zeros}). To study them, we set $k=i|k|$ in (\ref{zeros}) and write it in the form
    \be
    e^{-2|k|a}=\gamma\left(1+\frac{|k|}{\fz}\right).
    \label{bs1}
    \ee
According to this equation, a bound state exists if the right-hand side of this equation is a positive real number smaller than 1. For the case of Dirichlet and Neumann boundary conditions, where $\gamma=\pm 1$, this implies that $\fz$ must be a negative real number and $|k|<|\fz|$. For other values of $\gamma$, it is possible to use (\ref{bs1}) to obtain formulas for the phase angles of $\gamma$ and $\fz$ in terms of $|k|$, $|\fz|$, and $|\gamma|$. Because these are not illuminating, we do not include them here. We instead treat the problem of finding the conditions for the existence and uniqueness of the bound states for boundary conditions (\ref{bc}) with a real value of $\gamma$, which includes the physically relevant cases of the Dirichlet and Neumann boundary conditions. We differ the details of our analysis to the appendix and summarize its outcome in the following.
    	\begin{enumerate}
	\item There are at most two bound states.
	\item For the following two cases, no bound states exist.
		\begin{enumerate}
		\item $\gamma=0$;
		\item $1\geq\gamma\geq 2a|\fz|$.
		\end{enumerate}
	\item For the following three cases, there is a single bound state.
		\begin{enumerate}
		\item $\gamma<0$;
		\item $\gamma>1\geq 2a|\fz|/\gamma$;
		\item $1\leq\gamma<2a|\fz|$.
		\end{enumerate}
	\item There are two bound states if and only if 
	$0<\gamma<1< 2a|\fz|/\gamma$.
		\end{enumerate}
In particular for the Dirichlet and Neumann boundary conditions, there is at most a single bound state.  For the Dirichlet boundary condition, where $\gamma=1$, this is present if and only if $\fz$ is real and smaller than $-1/2a$. For the Neumann boundary condition, where $\gamma=-1$, a bound state exists and is unique if and only if $\fz$ is a negative real number.

In general, the computation of the value of $|k|$ for the bound states requires a numerical search for solutions of (\ref{bs1}). It is easy to see from this equation that solutions lie in the interval between $(1-|\gamma|^{-1})|\fz|$ and $(1+|\gamma|^{-1})|\fz|$ whenever they exist.

\subsection{Spectral singularities}

Spectral singularities are given by the real and positive values of $k$ that fulfill (\ref{zeros}). To explore them, we first write this equation in the form
    \bea
    &&e^{2ika}=\gamma\left(1-\frac{2ik}{\fz}\right).
    \label{eq2}
    \eea
If we evaluate the modulus of both sides of (\ref{eq2}) and note that $k$ is real and nonzero, we find a quadratic equation in $k$ with solutions
    \be
    k=\frac{1}{2}\left[\fz_i \pm \sqrt{|\fz/\gamma|^{2}-\fz_r^2}\right],
    \label{ss-k}
    \ee
where $\fz_r$ and $\fz_i$ stand for the real and imaginary parts of $\fz$, respectively.

The right-hand side of (\ref{ss-k}) is real, if and only if
    \be
    |\gamma|\leq\frac{|\fz|}{|\fz_r|}=\sqrt{1+\frac{\fz_i^2}{\fz_r^2}}.
    \label{condi-ss1}
    \ee
This is clearly satisfied by both the Dirichlet ($\gamma=1$) and Neumann ($\gamma=-1$) boundary conditions. These are special cases of the boundary conditions characterized by $|\gamma|=1$ which also fulfill (\ref{condi-ss1}). For the latter, (\ref{ss-k}) yields a single positive and real value for $k$, namely
    \be
    k=\fz_i,
    \label{k-ss-spec}
    \ee
provided that $\fz_i>0$. Substituting (\ref{k-ss-spec}) in (\ref{eq2}) and solving for $\fz_r$, we find
    \be
    \fz_r=-\cot(a\,\fz_i-\varphi/2)\fz_i,
    \label{zr-ss-spec}
    \ee
where $\varphi$ is the principal argument (phase angle) of $\gamma$, so that $\gamma=e^{i\varphi}$.

In terms of the parameters $\alpha$ and $\beta$ that enter the expression for the boundary condition (\ref{bc}), $|\gamma|=1$ takes the form $\alpha^*\beta\in\R$. Under this condition, the $\delta$-function potential defined in the half-line admits a single spectral singularity, if $\fz_i>0$ and $\fz_r$ satisfies
(\ref{zr-ss-spec}). These conditions have a rather interesting physical meaning. To see this, we use the correspondence between the Schr\"odinger equation (\ref{sch-eq}) and the Helmholtz equation for a transverse electric wave that is normally incident upon a dielectric media with planar symmetry along the $y$- and $z$-directions \cite{pra-2011a}. For the problem at hand with Dirichlet boundary conditions $(\beta=\alpha-1=\gamma-1=0$), the medium consists of a thin planar slab made of high-gain material that is placed at a distance $a$ to the right of a perfect planar mirror. We identify the latter with the plane $x=0$. Assuming that the system is placed in vacuum, we can model its permittivity profile using
    \be
    \varepsilon(x)=\varepsilon_0[1+\zeta\,\delta(x-a)],
    \label{eps=}
    \ee
where $x\geq 0$, $\varepsilon_0$ is the permittivity of the vacuum, and $\zeta$ is a complex coupling constant. The Helmholtz equation describing the behavior of normally incident transverse electric waves takes the form of the Schr\"odinger equation (\ref{sch-eq}) provided that the potential $v(x)$ is given by (\ref{delta-half}) and
    \be
    \fz=-k^2\zeta.
    \label{z-z}
    \ee
The presence of a spectral singularity for a wavenumber $k$ means that the above system fulfills the so-called laser threshold condition \cite{silfvast}; it operates as a slab laser whenever its gain exceeds the value associated with the spectral singularity.

For a realistic thin slab, we identify the $\delta$-function appearing in (\ref{eps=}) with a rectangular barrier potential with width $b\ll k^{-1}$ and complex hight $\zeta/b$. In this way we can identify the complex relative permittivity of the slab (which is equal to the square of its refractive index) with
    \be
    \hat\varepsilon_s:=1+\frac{\zeta}{b}=1-\frac{\fz}{b k^2}.
    \label{eps-slab}
    \ee
It is well-known that $\IM(\hat\varepsilon_s)>0$ and $\IM(\hat\varepsilon_s)<0$ respectively correspond to lossy and active optical material. In fact, the gain coefficient $g$ of the slab is related to $\IM(\hat\varepsilon_s)$ according to \cite{silfvast}
    \be
    g=-\frac{k\,\IM(\hat\varepsilon_s)}{\RE(\hat\varepsilon_s)}.
    \label{g=}
    \ee
Substituting (\ref{eps-slab}) in (\ref{g=}), we have
    \be
    g=\frac{\fz_i}{bk-\fz_rk^{-1}}.
    \label{g=1}
    \ee
In view (\ref{eps-slab}) and (\ref{g=1}), and the fact that for nonexotic active material, $\RE(\hat\varepsilon_s)>0$, the condition $\fz_i>0$ is equivalent to $g>0$. This means that the presence of a spectral singularity, which is the condition for the initiation of laser oscillations, requires the slab to be made of gain material; a well-known fact that is usually justified using the principle of conservation of energy.

For a perfect mirror, $\gamma=1$, and the spectral singularity occurs for $k=\fz_i$. This together with (\ref{g=1}) identifies the threshold gain for this slab laser with
    \be
    g=\frac{1}{b\, \RE(\hat\varepsilon_s)}.
    \label{g=b}
    \ee
Because $\RE(\hat\varepsilon_s)\geq 1$ and $bk\ll 1$, for wavelengths of the order of a micrometer, we have $g\gg 10^5~{\rm cm}^{-1}$, which is an extremely large gain coefficient. Notice however that this is to be produced in an extremely thin slab of thickness $b\ll k^{-1}\approx 1\,\mu{\rm m}$.

Next, we recall that to initiate lasing, we also need to satisfy (\ref{zr-ss-spec}). Because $\gamma=1$, we set $\varphi=0$ in this equation and use (\ref{eps-slab}) to write it in the form
    \be
    \RE(\hat\varepsilon_s)=1+\frac{\cot(ak)}{bk}.
    \label{eq201}
    \ee
Solving this equation for $a$, we find
    \be
    a=k^{-1}\left[-{\rm arctan}\left\{bk[\RE(\hat\varepsilon_s)-1]\right\}+
    \pi (m+\frac{1}{2})\right],
    \label{a=}
    \ee
where $m=0,1,2,\cdots$ is a mode number.

Given that $\RE(\hat\varepsilon_s)$ is at most of the order of $10$ and $bk\ll 1$, (\ref{a=}) implies
    \be
    a\approx \frac{(2m+1)\pi}{2k}=\frac{(2m+1)\lambda}{4},
    \label{a=app}
    \ee
where $\lambda:=2\pi/k$ is the wavelength. This in turn gives the following expression for the lasing modes of our thin-slab laser.
    \be
    k_m\approx\frac{(2m+1)\pi}{2a}.
    \label{modes}
    \ee
The particular mode $k_{m_\star}$ at which the laser will operate depends on the details of the dispersion relation describing the $k$-dependence of $\hat\varepsilon_s$. Making the $k$-dependence of $\IM(\hat\varepsilon_s)$ explicit and employing (\ref{g=}) and (\ref{g=b}), we can identify ${m_\star}$ with the mode number $m$ for which
    $\left|bk_{m}\IM[\hat\varepsilon_s(k_{m})]+1\right|$
takes its smallest possible value.

Notice that to find $m_\star$, we need to fix a particular value for $a$, which we denote by $a_0$. By construction, for $a=a_0$ and $k=k_{m_\star}$, Eq.~(\ref{eq201}) holds approximately. To ensure that it holds exactly, we make the $k$-dependence of $\RE(\hat\varepsilon_s)$ explicit and set $k=(2m_\star+1)\pi/2a$, so that (\ref{eq201}) turns into a single real equation which we can in principle solve for $a$. For this value of $a$ which we label by $a_\star$, our thin-slab laser will emit coherent waves with wavelength $\lambda_{m_\star}=4a_\star/(2m_\star+1)$ as soon as its gain coefficient  surpasses its threshold value, namely $b^{-1}$.

The requirement that the $\delta$-function potential in the half-line with Dirichlet boundary conditions at $x=0$ realizes a spectral singularity does not fix the real part of its coupling constant. This is in sharp contrast with the case of a $\delta$-function potential in the full line, i.e., the situation where we remove the mirror in our thin-slab system. In this case, the spectral singularity appears for purely imaginary values of the coupling constant \cite{jpa-2006b}. This corresponds to a high-gain thin slab with $\RE(\hat\varepsilon)=1$ which is extremely difficult to realize, because the real part of the permittivity of the known high-gain material is larger than unity. This signifies a practical advantage of the setup modeled by the $\delta$-function potential in the half-line.

\subsection{Perfect absorption}

The $\delta$-function potential in the half-line can serve as a perfect absorber provided that (\ref{CPA}) holds. We can write this equation in the form
    \be
    e^{2ika}=\frac{\gamma\fz}{\fz+2ik},
    \label{pa-delta}
    \ee
and use the fact that we are interested in real and positive values of $k$ satisfying this equation to show that it implies
    \be
    k=\frac{1}{2}\left[-\fz_i\pm\sqrt{|\gamma\,\fz|^2-\fz_r}\right].
    \label{pa-1}
    \ee
Because $k$ is real,
    \be
    |\gamma|\geq\frac{|\fz_r|}{|\fz|}=\frac{1}{\sqrt{1+\fz_i^2/\fz_r^2}}.
    \label{pa-condi}
    \ee

For the class of boundary condition specified by $|\gamma|=1$, (\ref{pa-condi}) holds and (\ref{pa-1}) gives a single positive value for $k$, namely
    \be
    k=-\fz_i,
    \label{pa-k=}
    \ee
whenever $\fz_i<0$. Again in the context of the optical model we discussed in the preceding subsection, this condition indicates that the system can act as a perfect absorber, if the thin slab is made of lossy material. Notice again that this is just a necessary condition for the perfect absorption of waves with wavenumber $k=-\fz_i$. The wave will be absorbed if (\ref{pa-delta}) holds for this value of $k$. This happens for
    \be
    \fz_r=-\cot(a\fz_i+\varphi/2)\fz_i=-\cot(a|\fz_i|-\varphi/2)|\fz_i|.
    \label{pa-zr}
    \ee
Again we can use a procedure similar to the one we described in the preceding subsection to determine the position $a_\star$ of the slab at which it acts as a perfect absorber for a wave with wavelength $\lambda_\star$. This involves using (\ref{a=}) and (\ref{a=app}) with $\fz_i$ replaced with $|\fz_i|$. These show that for given values of $\varphi$ and $\fz_r$, we can make the slab act as a perfect absorber provided that we adjust its position properly.

Equations (\ref{pa-k=}) and (\ref{pa-zr}) are consistent with the fact that the potential acts as a perfect absorber if and only if its time-reversal realizes a spectral singularity. To see this, we observe that under time-reversal transformation, $\fz_r\to\fz_r$, $\fz_i\to-\fz_i$, $\alpha\to\alpha^*$, $\beta\to\beta^*$, and consequently $\gamma\to 1/\gamma^*$. In particular for $|\gamma|=1$, we can determine the configurations of the system that make it act as a perfect absorber from those giving rise to a spectral singularity by changing $\fz_i\to-\fz_i$. This is consistent with the fact that performing this transformation in (\ref{k-ss-spec}) and (\ref{zr-ss-spec}) we respectively recover  (\ref{pa-k=}) and (\ref{pa-zr}).

\section{A slab laser with one mirror}

Consider a laser obtained by placing an infinite planar homogeneous slab of gain material with thickness $L$ next to a nearly perfect mirror \cite{silfvast}. We wish to determine how the threshold gain for a normally incident transverse electric wave depends on the reflectivity of the mirror and its distance to the slab. We do this by exploring the spectral singularities of the scattering problem for an optical potential $v(x)$ defined on the half-line, which represents the interaction of the wave with the slab, and a boundary condition at $x=0$ that signifies the effect of the mirror. 

The optical potential representing the slab has the form
	\be
	v(x):=\left\{\begin{array}{cc}
	k^2(1-\fn^2) &{\rm for}~ x\in[a,a+L],\\
	0& {\rm otherwise},\end{array}\right.
	\label{pot-sys}
	\ee
where $\fn$ is the complex refractive index of the slab, and $a$ is its distance to the mirror. The choice of the boundary condition at $x=0$ is dictated by the fact that we can identify the reflection amplitude of a nearly perfect mirror with $\cR_-=-1+\epsilon$, where $\epsilon$ is a complex number such that $|\epsilon|\ll 1$. In view of (\ref{bc}), (\ref{gamma=}), and the analysis leading to (\ref{bc-zero}), this corresponds to setting $\alpha=i(\cR_--1)=-i(2-\epsilon)$, $\beta=\cR_-+1=\epsilon$, and 
	\be
	\gamma=1-\epsilon.
	\label{gamma=1}
	\ee 
	
To determine the laser threshold condition, we require that the above system has a spectral singularity \cite{pra-2011a}, i.e., (\ref{ss}) holds. To explore consequences of this equation, we use (\ref{RRT=}) and the following well-known formula for the transfer matrix of the potential (\ref{pot-sys}) to determine $R^r$ for this potential.
	\be
	\bM=\left[\begin{array}{cc}
    	[\cos(kL\fn)+i\fn_+\sin(kL\fn)]e^{-ikL} &
    	i\fn_-\sin(kL\fn)e^{-ik(L+2a)}\\[6pt]
    	-i\fn_-\sin(kL\fn)e^{ik(L+2a)} &
    	[\cos(kL\fn)-i\fn_+\sin(kL\fn)]e^{ikL}\end{array}\right],
    	\label{am-barrier-M}
    	\ee
where $\fn_\pm:=(\fn\pm\fn^{-1})/2$, \cite{book-chapter}. This yields, 
	\[R^r=\frac{i\fn_-e^{-2ik(L+a)}}{\cot(kL\fn)-i\fn_+}.\]
Substituting this equation in (\ref{ss}) and using various properties of trigonometric functions, we can express the latter in the form:
	\bea
	e^{2ikL\fn}&=&
	\frac{{\tilde\fn}(1+{\tilde\fn} X)}{X+{\tilde\fn}},
	\label{SS-silver-2a}
	\eea
where
	\begin{align}
	&{\tilde\fn}:=\frac{\fn+1}{\fn-1},
	&&X:=\gamma\, e^{2ik(a+L)}=(1-\epsilon)e^{2ik(a+L)}.
	\label{tilde-n}
	\end{align}

Next, we recall that the gain coefficient $g$ for the slab is related to the imaginary part $\kappa$ of the refractive index $\fn$ according to $g=-4\pi{\kappa}/\lambda$, \cite{silfvast}. Expressing $\kappa$ in terms of $g$ and denoting the real part of $\fn$ by $\eta$, so that
	\be
	\fn=\eta+i\kappa=\eta-\frac{i\lambda g}{4\pi}=
	{\eta}-\frac{ig}{2k},
	\label{n1=}
	\ee
we can write (\ref{n1=}) as
	\be
	e^{gL}=\frac{{\tilde\fn}(1+{\tilde\fn} X) e^{-2ikL{\eta}}}{X+{\tilde\fn}}.
	\label{SS-silver-2}
	\ee
Because $e^{gL}$ is real and positive, we can equate it to the modulus of the right-hand side of (\ref{SS-silver-2}). This gives the following expression for the threshold gain.
	\be
	g=g^{(s)}+g^{(m)},
	\label{g=gg}
	\ee
where
	\begin{align}
	&g^{(s)}:=\frac{2}{L}\ln\left|{\tilde\fn}\right|,
	\label{gg=s}\\
	&g^{(m)}:=
	\frac{1}{L}\ln\left|\frac{X+{\tilde\fn}^{-1}}{X+{\tilde\fn}}\right|.
	\label{gg=m}
	\end{align}
Note that the right-hand side of (\ref{gg=s}) coincides with the known formula for the threshold gain of a mirrorless slab laser \cite{pra-2011a,silfvast}. Therefore, $g^{(m)}$ represents the contribution of the mirror. 

Next, we use the fact that the right-hand side of  (\ref{SS-silver-2}) is real to infer,	\be
	k=\frac{2\pi m+\vartheta}{2\eta L},
	\label{laser-modes}
	\ee
where $m=0,1,2,3,\cdots$ is a mode number,  and $\vartheta$ is the principal argument (phase angle) of 
${\tilde\fn}(1+{\tilde\fn} X)/(X+{\tilde\fn})$, i.e., the real number belonging to $[0,2\pi)$ that satisfies
	\be
	e^{i\vartheta}=\frac{|X+\tilde\fn|\tilde\fn(1+\tilde\fn X)}{|\tilde\fn(1+\tilde\fn X)|(X+\tilde\fn)}.
	\label{mode-angle}
	\ee
Eq.~(\ref{laser-modes}) gives the lasing modes of our slab laser. 
	
For a vast majority of typical active material, $|{\kappa}|\ll {\eta}-1$, and we can safely ignore the $\kappa$-dependence of the right-hand sides of (\ref{gg=s}), (\ref{gg=m}), and (\ref{mode-angle}). This gives
	\begin{align}
	g^{(s)}=&\frac{2}{L}\ln\left(\frac{{\eta}+1}{{\eta}-1}\right)+\cO({\kappa}^2),
	\label{app1}\\
	g^{(m)}=&\frac{1}{L}\left[\ln\left(\frac{{\eta}-1}{{\eta}+1}\right)+
	\ln\left|Y\right|\right]+
	\cO({\kappa}),
	\label{app2}\\
	e^{i\vartheta}=&\frac{Y}{|Y|}+
	\cO({\kappa}),
	\label{app3}
	\end{align}
where 
	\[ Y:=\frac{({\eta}+1) X+{\eta}-1}{({\eta}-1)X+{\eta}+1},\]
and
$\cO(\kappa^\ell)$ stands for the sum of terms of order $\ell$ or higher in powers of $\kappa$. We can obtain the following more explicit expressions for the right-hand sides of (\ref{app2}) and (\ref{app3}), if we employ the definition of $X$, i.e., (\ref{tilde-n}), and the fact that $|\epsilon|\ll 1$. 
	\bea
	g^{(m)}&=&-\frac{1}{L}\left[\ln\left(\frac{{\eta}+1}{{\eta}-1}\right)
	+2Z\,\RE(\epsilon)\right]+\cO({\kappa})+\cO(\epsilon^2),
	\label{gm=0}\\
	\vartheta&=&\vartheta_s-2Z\,\IM(\epsilon)+\cO({\kappa})+\cO(\epsilon^2),
	\eea
where
	\bea
	Z&:=&\frac{\eta}{{\eta}^2+1+({\eta}^2-1)\cos[2k(a+L)]}.\\
	\vartheta_s&:=&2\,{\rm arctan}\left\{\eta^{-1}\tan[k(L+a)]\right\}.
	\eea
	
The fact that the right-hand side of (\ref{gm=0}) takes a negative value is consistent with the expectation that placing the slab next to a mirror reduces its threshold gain. What is not expected is that a realistic mirror whose reflection amplitude has a slightly larger real part than $-1$, so that $\RE(\epsilon)=1+\RE(\cR_-)>0$, leads to a larger reduction of the threshold gain in comparison to a perfect mirror! Eqs.~(\ref{g=gg}), (\ref{app1}), and (\ref{gm=0}) lead to the following formula for the threshold gain.
	\be
	g=\frac{1}{L}\left\{\ln\left(\frac{{\eta}+1}{{\eta}-1}\right)
	-\frac{2{\eta}\,\RE(\epsilon)}{{\eta}^2+1+({\eta}^2-1)\cos[2k(a+L)]}\right\}
	+\cO({\kappa})+\cO(\epsilon^2),
	\label{g=0}
	\ee
According to this equation placing a gain slab next to a perfect mirror reduces its threshold gain by a factor of 2. This agrees with the fact the presence of a perfect mirror has the same effect as doubling the thickness of the slab (the mirror image of the slab acts as a second amplifier \cite{silfvast}.) Another consequence of Eq.~(\ref{g=0}) is that the threshold gain takes its smallest value, namely
	\[g_{\rm min}=\frac{1}{L}\left\{\ln\left(\frac{{\eta}+1}{{\eta}-1}\right)
	-\frac{2{\eta}\,\RE(\epsilon)}{{\eta}^2+1}\right\}
	+\cO({\kappa})+\cO(\epsilon^2),\]
   whenever
   	\[ a+L=\frac{(2\ell+1)\pi}{4k}=\frac{(2\ell+1)\lambda}{8},\]
	for some nonnegative integer $\ell$. This relation identifies the optimal positions of the slab. Again it is interesting to observe that for a perfect mirror the distance between the slab and the mirror does not affect the threshold gain, while for a nearly perfect mirror one can lower the threshold gain by adjusting the distance.

\section{Extension to short-range nonlinear scatterers}

Consider the scattering problem defined on the half-line by the a nonlinear Schr\"odinger equation of the form,
	\be
	-\psi''(x)+\cF(\psi(x),x)=k^2\psi(x),
	\label{NL-sch-eq}
	\ee
and the boundary condition (\ref{bc}), where $\cF:\C\times[0,\infty)\to\C$ is a function such that as $x\to\infty$, $\cF(\psi(x),x)$ tends to $v(x)\psi(x)$ for some scattering potential $v:[0,\infty)\to\C$. Ref.~\cite{epjp-2019} provides a detailed discussion of the scattering problem defined by (\ref{NL-sch-eq}) in the full line. In particular, it introduces a nonlinear generalization of the transfer matrix which shares the basic properties of its well-known linear predecessor. In this section, we employ the analysis of Sec.~2 to treat the scattering problem given by (\ref{bc}) and (\ref{NL-sch-eq}) in the half-line. 

First, we recall that the nonlinear transfer matrix is a complex $2\times 2$ matrix $\bM$ satisfying (\ref{M-def}) that depends on the amplitudes $A_-$ and $B_-$ (in addition to the wavenumber $k$.) We therefore denote it by $\bM(A_-,B_-)$. The left/right reflection and transmission amplitudes, $R^{l/r}$ and $T^{l/r}$, for the scattering problem given by (\ref{NL-sch-eq}) in the full line are also defined by (\ref{RT-def}). But, in general, they depend on the amplitude of the incident wave. Furthermore, the presence of nonlinearity can violate the reciprocity in transmission, i.e., for identical incident waves $T^l$ and $T^r$ are generally different. 

The formulas relating $R^{l/r}$ and $T^{l/r}$ to the nonlinear transfer matrix $\bM(A_-,B_-)$ are the following analogs of (\ref{RRT=}).
	\be
    \begin{aligned}
    &R^l=-\frac{M^l_{21}}{M^l_{22}},\hspace{1cm}
    &&T^l=\frac{\det(\bM^l)}{M^l_{22}},\\
    &R^r=\frac{M^r_{12}}{M^r_{22}},\hspace{1cm}
    &&T^r=\frac{1}{M^r_{22}},
    \end{aligned}
    \label{NL-RRT=}
    	\ee
where $M^{l}_{ij}$ and $M^{r}_{ij}$ are respectively the entries of 
	\begin{align}
	&\bM^{l}:=\bM(A^l,A^lR^l),
	&&\bM^{r}:=\bM(0,A^rT^r),
	\label{NL-M=}
	\end{align}
and $A^{l/r}$ is the complex amplitude of the left/right incident wave, i.e., $A^l=A_-$ when $B_+=0$, and $A_+=B_-$ when $A_-=0$, \cite{epjp-2019}. In view of (\ref{NL-M=}), we can use (\ref{NL-RRT=}) to express $R^{l/r}$ and $T^{l/r}$ in terms of $A^{l/r}$.

An important distinction between linear and nonlinear transfer matrices is that Eq.~(\ref{M-def}) determines the latter up to a pair of unspecified functions $f_1(A_-,B_-)$ and $f_2(A_-,B_-)$; if $\bM(A_-,B_-)$ satisfies (\ref{M-def}), then so does 
	\[\tilde \bM(A_-,B_-):=\bM(A_-,B_-)+\left[\begin{array}{cc}
	f_1(A_-,B_-)B_-~~&~~-f_1(A_-,B_-)A_-\\
	f_2(A_-,B_-)B_-~~&~~-f_2(A_-,B_-)A_-\end{array}\right].\]
For the linear transfer matrix, one avoids this problem by demanding that it is independent of $A_-$ and $B_-$.  For the nonlinear transfer matrix this requirement is never fulfilled, and there is no general guideline to make a particular choice for $f_1(A_-,B_-)$ and $f_2(A_-,B_-)$. Fortunately, this large lack of uniqueness does not affect the utility of the nonlinear transfer matrix in determining the reflection and transmission amplitudes, for it happens that $\bM(A_-,B_-)$ and $\tilde \bM(A_-,B_-)$ give rise to the same formulas for the reflection and transmission amplitudes, \cite{epjp-2019}. 

Now, consider the scattering problem given by (\ref{NL-sch-eq}) and (\ref{bc}) on the half-line. Solving this problem means finding the reflection amplitude $\cR$ defined by (\ref{ref-coeff}). Noting that the source for the incident wave is located at $x=+\infty$, $B_+=A^r$, and (\ref{ref-coeff}) reads 
	\be
	\cR:=\frac{A_+}{A^r}.
	\label{NL-R-def0}
	\ee 
Repeating the analysis of Sec.~2 for our nonlinear scattering problem, we are led to Eqs.~(\ref{q101}) and (\ref{q102}) with $M_{ij}$ replaced with the entries $ \cM_{ij}(A_-)$ of 
	\be
	\bcM(A_-):=\bM(A_-,-\gamma\,A_-).
	\label{bcM-def}
	\ee
Substituting $\cM_{ij}(A_-)$ for $M_{ij}$ in (\ref{q101}) and (\ref{q102}), we arrive at a pair of equations that are respectively equivalent to
	 \bea
    	&&\cR=\frac{A_-\left[\cM_{11}(A_-)-\gamma \cM_{12}(A_-)\right]}{A^r},
    	\label{NL-q101}\\[6pt]
    	&&\left[\cM_{21}(A_-)-\gamma \cM_{22}(A_-)\right]A_-=A^r.
   	 \label{NL-q102}
    	\eea
It is not difficult to check that the above-mentioned lack of uniqueness of the nonlinear transfer matrix does not affect these equations either; in view of (\ref{bcM-def}), the transformation $\bM(A_-,B_-)\to\tilde\bM(A_-,B_-)$ leaves $\cM_{11}(A_-)-\gamma \cM_{12}(A_-)$ and $\cM_{21}(A_-)-\gamma \cM_{22}(A_-)$ invariant.

If we know a nonlinear transfer matrix for the scattering problem defined in the full line, we can in principle solve (\ref{NL-q102}) for $A_-$ and substitute the result in (\ref{NL-q102}) to determine $\cR$ as a function of $A^r$, i.e., solve the corresponding nonlinear scattering problem in the half-line. It is also easy to see that the bound states (respectively spectral singularities) are given by the nonzero values of the complex amplitude $A_-$ and positive imaginary (respectively positive real) values of $k$ for which
	\bea
	\cM_{11}(A_-)&\neq&\gamma\, \cM_{12}(A_-),
	\label{NL-SS1}\\
	\cM_{21}(A_-)&=&\gamma\, \cM_{22}(A_-).
	\label{NL-SS2}
	\eea
Note that in this case,
	\be
	A_+=A_-\left[\cM_{11}(A_-)-\gamma\, \cM_{12}(A_-)\right]\neq 0,
	\label{NL-SS3}
	\ee
while $B_+=A^r=0$.

As a simple example, consider the scattering problem defined by the nonlinear point interaction given by
	\be
	\cF(\psi(x),x):=f(|\psi(x)|)\,\psi(x)\,\delta(x-a),
	\label{NL-delta1}
	\ee
where $a$ is positive real parameter, and $f:[0,\infty)\to\C$ is a continuous function. As shown in Ref.~\cite{epjp-2019}, we can obtain a nonlinear transfer matrix $\bM(A_-,B_-)$ associated with (\ref{NL-delta1}) by identifying the coupling constant $\fz$ of its linear analog, namely (\ref{delta-half}), with $f(|e^{2ic\fK}A_-+B_-|)$. In other words, $M_{ij}(A_-,B_-)$ are given by (\ref{Mij=}), if we set $\tilde\fz:=f(|e^{2iak}A_-+B_-|)/2k$. In particular, for $B_-=-\gamma\,A_-$,
	\be
    	\tilde\fz=\frac{f(|\tilde\gamma\,A_-|)}{2k}=\frac{f(|A_+|)}{2k},
    	\label{NL-tz}
    	\ee
where 
	\be
	\tilde\gamma:=\gamma-e^{2iak},
	\label{t-gamma}
	\ee
and we have made use of the identity,
	\be
	A_+=e^{-2iak}\tilde\gamma A_-,
	\label{A-plus=}
	\ee 
which follows from (\ref{Mij=}), (\ref{NL-SS3}), and (\ref{NL-tz}).

In view of (\ref{Mij=}), (\ref{NL-tz}), and (\ref{t-gamma}), (\ref{NL-q102}) takes the form
	\be
	\left[\tilde\gamma f(|\tilde\gamma\,A_-|)-2ik
	\left(\tilde\gamma+e^{2iak}\right)\right]A_-=2ik A^r.
	\label{NL-q102-2}
	\ee
Evaluating the modulus of both sides of this equation, we see that $|A_-|$ is a function of $|A^r|$. This together with (\ref{NL-q102-2}) allow us to determine the phase angle for $A_-$. Similarly, we use (\ref{Mij=}), (\ref{NL-tz}), and (\ref{t-gamma}), to simplify the expression given by (\ref{NL-q101}) for $\cR$. This yields
	\be
	\cR= \frac{A_-(1+i\tilde\gamma\,\tilde\fz\, e^{-2iak})}{A^r}
	=-e^{-2iak}\left(1+\frac{\tilde\gamma A_-}{A^r}\right),
	\label{NL-R-delta}
	\ee
where we have also employed (\ref{NL-q102-2}).

The nonlinear point interaction (\ref{NL-delta1}) admits a (nonlinear) spectral singularity \cite{prl-2013} provided that (\ref{NL-SS1}) and (\ref{NL-SS2}) holds for some positive real $k$ and $A_-\neq 0$. In view of (\ref{t-gamma}), the second of these equations implies
	\be
	f(|A_+|)=
	2ik\left(1-\frac{e^{2iak}}{\gamma}\right)^{-1}.
	\label{NL-R-delta2}
	\ee
This is a particularly useful formula, for it relates $|A_+|$, which determines the intensity of the outgoing wave, to the wavenumber $k$ and the parameters entering the expression for the nonlinearity profile $f(|\psi|)$.

Next, we recall the application of the $\delta$-function potential for modeling the scattering of a TE wave by an active thin slab that is placed next to a perfect mirror ($\gamma=1$). We may identify the scattering problem for this potential with the one defined by (\ref{NL-delta1}), if we set $f(|\psi(x)|)=\fz$. Replacing this equation with
	\be
	f(|\psi(x)|)=\fz+\fs\,|\psi(x)|^2,
	\label{kerr}
	\ee
corresponds to a Kerr slab with relative permittivity (\ref{eps-slab}), Kerr constant 
	\be
	\sigma:=-\frac{\fs}{b k^2},
	\label{kerr-slab}
	\ee
and thickness $b\ll k$. If $k^2$ is a spectral singularity associated with the nonlinearity profile (\ref{kerr}), the slab will function as a laser with a single mirror and resonance wavelength $k$. In this case, we can use (\ref{NL-R-delta2}) to express the intensity of the outgoing laser light to the real part of the relative permittivity of the slab and its gain coefficient. 

Substituting (\ref{kerr}) in (\ref{NL-R-delta2}) and employing the analysis leading to (\ref{g=b}) and (\ref{eq201}), we obtain the following nonlinear analogs of these equations.
	\bea
	&&g=g_0[1+2 bk\,\IM(\sigma)I],
	\label{NL-g=b}\\
	&&\RE(\hat\varepsilon_s)+2\,\RE(\sigma)I=1+\frac{\cot(ak)}{bk},
	\label{NL-eq201}
	\eea
where $g_0:=1/b\,\RE(\hat\varepsilon_s)$, and $I:=|A_+|^2/2$ is the time-averaged intensity of the outgoing wave. According to (\ref{NL-g=b}) a spectral singularity is realized at $g=g_0$ when $I=0$. This suggests identifying $g_0$ with the threshold gain whenever $\IM(\sigma)>0$. 

Suppose that the source of nonlinearity is the nonlinear response of the gain material, and assume that it deduces the amplification of waves propagating inside the slab. Then $\IM(\sigma)$ must indeed take a positive value. This observation together with the fact that we can express (\ref{NL-g=b}) in the form
	\be
	I=\frac{g-g_0}{2bk\,\IM(\sigma)},
	\label{I=33}
	\ee
 show that in order for the slab to emit laser light its gain coefficient must exceed $g_0$. This identifies $g_0$ with the threshold gain. Moreover, it implies that for gain values exceeding $g_0$, the intensity of the outgoing laser light is proportional to $g-g_0$. This is in complete agreement with established facts about lasers.
 
Notice also that, according to (\ref{NL-eq201}), whenever the real part of the Kerr coefficient is nonzero, the wavenumber of the emitted wave undergoes a shift. Substituting (\ref{I=33}) in (\ref{NL-eq201}), we have
	\be
	\cot(ak)-[\RE(\hat\varepsilon_s)-1]bk=\frac{(g-g_0)\RE(\sigma)}{\IM(\sigma)}.
	\label{eq471}
	\ee
To reach threshold $g=g_0$, we must adjust the distance $a$ between the slab and the mirror so that the left-hand side of this relation vanishes for the wavenumber $k$ for which we can maintain the necessary gain. Once we increase the gain to obtain a positive intensity for the outgoing wave, the right-hand side becomes positive and as a result the value of $k$ changes. Because $bk\ll 1$, (\ref{I=33}) shows that the right-hand side of (\ref{eq471}) is extremely small. This in turn allows us to determine the shift $\delta k$ in the wavenumber using first-order perturbation theory. The result is
	\[\frac{\delta k}{k}\approx
	-\frac{(g-g_0)\RE(\sigma)}{\{a+b[\RE(\hat\varepsilon_s)-1]\}k\,\IM(\sigma)}	
	\approx-\frac{(g-g_0)\RE(\sigma)}{ak\,\IM(\sigma)}=
	-\frac{2b\RE(\sigma)I}{a},\]
where we have benefitted from the fact that $bk\ll 1$ and $b\ll a$.

\section{Concluding remarks}

In this article, we have derived a simple formula for the reflection amplitude of a general class of scattering problems on the half-line in terms of the reflection and transmission amplitudes of a scattering potential defined on the full line. This provides the means for using the methods developed to solve the scattering problems in the full line to deal with those in the half-line. 

The potential scattering in the half-line $[0,\infty)$ involves a boundary condition at $x=0$ which we have taken to be of the form (\ref{bc}). This is of particular interest because it appears in the treatment of the scattering problem for the extensions to the whole line of a potential $v(x)$ defined on the half line. Suppose that $v:[0,\infty)\to\C$ and $v_-:(-\infty,0)\to\C$ are scattering potentials and there is some $\epsilon>0$ such that $v(x)=0$ for $x<\epsilon$. Then we can employ the argument leading to (\ref{psi-zero}) to identify the right-reflection amplitude $\cR$ of the potential
    \[V(x):=\left\{\begin{array}{ccc}
    v_-(x) & {\rm for} & x<0,\\
    v(x) & {\rm for} & x\geq 0,
    \end{array}\right.\]
with the reflection amplitude of $v(x)$ provided that we impose the boundary condition (\ref{bc}) with
    \begin{align*}
    &\alpha=i(\cR_--1),
    &&\beta=\cR_-+1.
    \end{align*}
By making this choice for $\alpha$ and $\beta$ and using our approach to map the scattering problem for the potential $v(x)$ to that of its trivial extension to the full line, we can find the scattering properties of the potential $V(x)$. This argument shows that we can use our approach to determine the effect of placing a scatterer next to another one.

We have employed this approach to characterize the bound states, spectral singularities, and reflectionless (perfect absorbing) configurations of a scattering problem in the half-line in terms of certain conditions on the transfer matrix or reflection and transmission amplitudes of its trivial extension to the whole line. We demonstrated the utility of our general results in the study of $\delta$-function and barrier potentials defined on the half-line, and provided physical interpretation of the outcome in terms of simple optical realizations of these potentials.

Our approach admits a straightforward extension to short-range nonlinear scattering interactions. This is particularly useful because it allows for using nonlinear transfer matrices to study short-range nonlinear interactions that are defined in the half-line.

\section*{Appendix: Bound states of the $\delta$-function potential (\ref{delta-half}) for real $\gamma$} 

The bound states of $\delta$-function potential (\ref{delta-half}) that are determined by the boundary condition (\ref{bc}) are characterized by Eq.~(\ref{bs1}). For real values of $\gamma$, the requirement that the right-hand side of this equation is real, positive, and smaller than 1 shows that $\gamma\neq 0$ and $\fz$ must be real and negative, i.e., $\fz=-|\fz|\neq 0$. In light of this observation and Eq.~(\ref{bs1}), the bound states correspond to the real zeros of the function, $f(x):=1-x/|\fz|-e^{-2ax}/\gamma$. Clearly, $f'(x)=2a e^{-2ax}/\gamma-1/|\fz|$ and $f''(x)=-4a e^{-2ax}/\gamma$. We use this information to characterize the number of bound states for the following cases separately.
    \begin{enumerate}
    \item $\gamma<0$: In this case, $f(0)>0>f'(x)$ for all $x>0$. This together with the fact that
        \be
        \lim _{x\to\infty}f(x)=-\infty,
        \label{lim}
        \ee
show that $f(x)$ has a single positive zero. Hence there is a unique bound state.
    \item $\gamma>0$: In this case $f'(x)=0$ if and only if $x=x_0:=\frac{1}{2a}\ln(2a|\fz|/\gamma)$. This shows that $f(x)$ has a single extremum point, namely $x_0$. Because $f''(x)<0$ for all $x\in\R$, this is a maximum point of $f(x)$. Now, consider the following possibilities.
        \begin{itemize}
        \item[2.1)] $2a|\fz|/\gamma\leq 1$: In this case, $x_0\leq 0$. Therefore, $f'(x)<0$ for $x>0$, i.e., $f(x)$ is a decreasing function in $[0,\infty)$.
            \begin{itemize}
            \item[2.1.a)] For $\gamma\leq 1$, $f(0)\leq 0$, and because $f(x)$ is a decreasing function in $[0,\infty)$, we have $f(x)<0$ for $x>0$. Consequently, there is no bound state.
            \item[2.1.b)]  For $\gamma>1$, $f(0)>0$ and because (\ref{lim}) holds, $f(x)$ has a unique positive zero. Therefore, the system has a unique bound state.
            \end{itemize}
        \item[2.2)] $2a|\fz|/\gamma> 1$: In this case, $x_0>0$, and we can write the maximum value of $f(x)$ in the form $f'(x_0)=1-g(2a|\fz|/\gamma)/\gamma$, where $g(x):=(1+\ln x)/x$. Clearly $g'(x)=-(\ln x)/x^2<0$ for $x>1$. This shows that $g(x)$ is a decreasing function for $x>1$. Because $g(1)=1$, we conclude that $g(x)<1$ for $x>1$. Now, consider the following subcases.
            \begin{itemize}
            \item[2.2.a)] For $\gamma< 1$, the fact that $g(x)<1$ for $x>1$ implies that $f(x_0)>0$. This together with $f(0)<0$ and (\ref{lim}) show that $f(x)$ has two real and positive zeros, i.e., the system has two bound states.
            \item[2.2.b)] For $\gamma\geq 1$, again $f(x_0)>0$ but $f(0)\geq 0$. Therefore, in light of (\ref{lim}), $f(x)$ has a single positive zero. This implies the existence of a unique bound state.
            \end{itemize}
        \end{itemize}
        \end{enumerate}
The above considerations together with the previously mentioned fact that for $\gamma=0$ no bound state can exist complete the characterization of the number bound states for real values of $\gamma$.

\subsection*{Acknowledgments}
We are grateful to Farhang Loran for reading the first draft of this article.
This work has been supported by the Scientific and Technological Research Council of Turkey (T\"UB\.{I}TAK) in the framework of the project no: 114F357, and by the Turkish Academy of Sciences (T\"UBA).

\ed